
magnification=1200
\baselineskip=21pt
\overfullrule=0pt

\null
\vskip.5cm

\centerline{\bf STOCHASTIC DYNAMICS OF COARSE-GRAINED QUANTUM FIELDS IN}
\centerline{\bf THE INFLATIONARY UNIVERSE}
\vskip1cm
\centerline{Milan Miji\'c\footnote*{E-mail address:
milan@moumee.calstatela.edu}}
\vskip.5cm
\centerline{\it Department of Physics and Astronomy}
\centerline{\it California State University, Los Angeles, CA 90032}
\vskip2cm
\centerline{\bf Abstract}

It is shown how coarse-graining of quantum field
theory in de Sitter space leads to the emergence of a classical
stochastic description as an effective theory in the infra-red regime.
The quantum state of the coarse-grained scalar field is found to be a highly
squeezed coherent state, whose center performs a random walk on a
bundle of classical trajectories.

\vskip1truein
\centerline{PACS index: 98.80.-k}

\vfil\eject

{\bf 1. Introduction}
\medskip
Quantum fluctuations of matter fields play a prominent role in
inflationary cosmology [1]. They lead to cosmological
density perturbations that may be responsible for the origin of
structures in the Universe [2] and may completely alter our
concepts about the past, the future, and the global structure of
spacetime [3].

The fundamental idea behind the whole picture is that of a highly classical
behavior of quantum fluctuations with wavelengths larger than the Hubble
radius. It is fair to say however, that this has been widely used, but not
widely understood. Classicality of long wavelength quantum fluctuations in
de Sitter space has been established by Guth and Pi [4] in a mode-by-mode
treatment. Some related works are those in Ref. 5.

The purpose of this study is to look at the emergence of classical features in
the inflationary universe from another angle, by
using stochastic approach due to Starobinsky [6]. Instead of considering
each long wavelength mode separately, now the main object of study is their
collection---the coarse-grained field---which is just the properly defined
inflaton field that appears in every inflationary model. My goal here is to
attempt to clarify in what sense and for what reasons the evolution of that
field appears classical and stochastic. Two simplest, exactly solvable models
will be used, a massless and a massive scalar quantum field on an exact de
Sitter background. I will not consider either the scalar field driven
inflationary phase or the evolution of metric
fluctuations, but some qualitative extensions in these directions are
obvious and self-understood.

It is important to stress that while noise and stochastic forces are often
introduced in a semi-phenomenological manner, in this case the genuine
quantum fluctuations give rise to an effectively classical stochastic process
in the infra-red regime. For this reason, one would  expect the paradigm of
stochastic inflation not only to establish foundation for the scenario
of structure formation in post-inflationary universe,
but also to play an important role in understanding
applications of quantum mechanics to the early Universe---what may be
called the ``measurement problem,''
or, rather, the emergence of nearly classical spacetime in quantum cosmology.

As a more immediate goal, this work should help to exhibit a stochastic
process as underlying the true physical nature of the inflationary
phase. The essential physical elements that lead to a classical stochastic
description are: an inflationary expansion, the de Sitter vacuum,
a scalar field mass that is zero, or at least small compared to the Hubble
parameter, and a coarse-graining scale larger than the Hubble radius.
In particular, I tried to provide some
insight to the role that mass of the scalar field plays in the
emergence or absence of a classical stochastic regime.

\vskip0.5truein

{\bf 2. Langevin equation?}
\medskip
The power of stochastic inflation rests on Starobinsky's derivation [6]
of an effective equation for a coarse-grained scalar field on de Siter
background which looks very much like the Langevin equation in the theory of
Brownian motion. For a free, massive scalar field on
spatially flat background the result was,
$$
\dot {\Phi}_L = - {{m^2} \over {3H_0}} \Phi_L + \xi ~~.
\eqno (2.1)$$
This equation is just an approximate form of the Klein-Gordon equation in de
Sitter space. It applies to that component of a quantum field $\Phi$ which
is coarse-grained in the real space, so that, $\Phi_L (\vec {x} + \Delta
\vec {x}, t) \approx \Phi_L (\vec {x}, t)$, whenever $a(t)|\Delta \vec {x} |
\geq l \equiv (\epsilon_S H_0)^{-1} \gg H_0^{-1}$.
The last term on the right hand side of (2.1) represents the effect of the
constant inflow of the short-wavelength modes.
The time evolution is considered in time-steps $\Delta t$ no smaller than the
Hubble time $H_0^{-1}$.

The problem is that as yet there are no satisfactory arguments
for affirmative answer to crucial question: is coarse-grained quantum field
$\Phi_L$ really classical? That is, is Starobinsky
equation (2.1) indeed a Langevin equation?

At first glance the answer does appear affirmative. The two equations are of
the same form. If taking the vacuum expectation value is understood as
averaging over the Brownian noise the correspondence is remarkable:
$\langle \xi \rangle$ vanishes, and $\langle \xi \xi \rangle$
has the right form to make
the dispersion $\langle \Phi_L^2 \rangle$ linearly growing in time
for the case of a massless field. This result have been obtained at first
directly through mode summation [7], and led Vilenkin [8] to use
the Brownian interpretation even before equation (2.1) was written down.
The corresponding result for a massive field is also well known [7], but its
stochastic interpretation is less clear and has not been discussed much. (See
however Ref. 9.)

It is not necessary to discuss here in detail the relationship
between the stochastic approach to inflation and the standard quantum field
theory on de Sitter space. Let us just state the
obvious points. One usually acts as if equation (2.1) translates
into similar, classical equation for some expectation value of the quantum
field $\Phi_L$. However, the vacuum expectation value of $\Phi_L$ is
manifestly zero, just like the vacuum expectation value of $\xi$.
This may look consistent with random walk picture, but one should remember
that the expectation value for the
position of the random walker is zero only after averaging
over the noise. Invoking r.m.s. doesn't help, as clearly we cannot
write the Langevin equation for a quantity which is identically zero.

Moreover, everything that has been done so far were
just manipulations with the Klein-Gordon equation in quantum field theory.
What one normally ends up calculating are vacuum expectation values.
It appears that Eq. (2.1) is nothing more than the effective quantum field
equation at late times since, as one can easily check, it merely provides the
same expectation values which follow in the late time limit of the full
Klein-Gordon equation. One demonstration for this will be presented in
Section 6. The restriction to the coarse-grained field corresponds to the
usual subtraction of the ultraviolet-divergence due to short modes.

These are the clear signals that there is something more to be understood
about the Langevin interpretation of equation (2.1).
The real problem is in casual interpretation of vacuum averages
as Brownian noise averages. In the usual random walk the
linear growth of dispersion is obtained after averaging over the values of
actual steps, each of whom may be \lq\lq result of the measurement\rq\rq;
while in our case, when taking the vacuum expectation value,
we are averaging over possible, but unmeasured values of a quantum field.
Expressed in a different way, in the classical random walk we are averaging
over all the incoherent, classical stochastic histories that walker might have
taken up to the time $t$; while when taking the vacuum expectation value of a
quantum field $\Phi_L$ we are averaging over all the interfering histories
ending at some fixed value of a scalar field at the final hypersurface.
This is, of course, the essence of the problem to describe the transition
from quantum to classical regime.

The point is that the random walk interpretation does not follow merely from
the coarse-graining of the field equation. We have to show explicitly
that the coarse-grained field is indeed nearly classical, or, in the
language of Gell-Mann and Hartle [10], that its histories decohere.
Without that, Equation (2.1) cannot be interpreted
as Langevin equation. The task here is to investigate and construct such
argument.

This problem may be viewed in several ways: as establishing a physical picture
for the stochastic inflation; as a way to provide the proof
for the classicality of long-wavelength perturbations in de Sitter space; or
as a first step towards the understanding of the emergence of nearly classical
universe in quantum cosmology.

\vskip0.5truein

{\bf 3. The Commutators}
\medskip
Stochastic approach has been developed originally in the Heisenberg picture,
so it is appropriate to look for signs of nearly classical behavior by
examining relevant commutators. The variables of interest are the
coarse-grained field,
$$
\Phi_L \equiv \int {d {\vec k} \over (2\pi)^3}~\theta (k_S - k)
\left ( a_{\vec k} \phi_{\vec k} + a_{\vec k}^{\dagger}
\phi_{\vec k}^{*} \right )        ,
\eqno (3.1)$$
the coarse-grained momentum,
$$
\pi_L \equiv a^3(t) \int {d {\vec k} \over (2\pi)^3}~\theta (k_S - k)
\left ( a_{\vec k} \dot \phi_{\vec k} + a_{\vec k}^{\dagger}
\dot \phi_{\vec k}^{*} \right )        ,
\eqno (3.2)$$
and the two force terms due to the inflow of short modes, $\xi_S$
and $\xi_m$:
$$
\xi = \xi_S + {\xi_m \over 3H_0} ~~;
\eqno (3.3a)$$
$$
\xi_S \equiv k_S (t) H_0 \int {d\vec {k} \over (2\pi)^3} \delta (k - k_s(t))
[a_{\vec k}\phi_{\vec k}(x) + a_{\vec k}^{\dagger}\phi_{\vec k}^*(x) ]~~;
\eqno (3.3b)$$
$$
\xi_m \equiv k_S (t) H_0 \int {d\vec {k} \over (2\pi)^3} \delta (k - k_s(t))
[a_{\vec k}\dot{\phi}_{\vec k}(x)
+ a_{\vec k}^{\dagger}\dot{\phi}_{\vec k}^*(x) ]~~;
\eqno (3.3c)$$
$H_0$ is the Hubble parameter, truly constant in this case.
$k_S(t)\equiv \epsilon_S a(t) H_0$
stands for the wave number of a mode whose physical wavelength
at the given time is equal to the coarse-graining scale $l \equiv (\epsilon_S
H_0)^{-1}$, with $0<\epsilon_S \ll 1$. For spatially flat RW background the
modes $\phi_{\vec k}(x)$ may be expanded as,
$\phi_{\vec k}(x) = v_k(t) \exp [i\vec {k} \cdot \vec {x}]$.

{}From the structure of the operators in Eq.'s (3.1-3.3) it immediately follows
that,
$$
[\Phi_L(\vec {x}_1,t), \xi_S(\vec {x}_2,t)] =  0,~~~~~
[\pi_L(\vec {x}_1,t), \xi_M(\vec {x}_2,t)] =  0.
\eqno (3.4a,b)$$

To evaluate finite commutators one has to specify the modes.
To make a point, I will consider here only a massless field.
Using conformal time $\eta$, the time-dependent modes are given as,
$v_k(\eta) = (H_0 /\sqrt{2 k})[\eta - i /k ]
\exp [ (- i)(k\eta - \pi/4)]$.
Introducing $R \equiv |\vec {x}_1 - \vec {x}_2|$,
one finds,
$$
[\Phi_L(\vec{x}_1, t), \pi_L(\vec{x}_2, t)]
={i \over 2\pi^2}{k_S \over R^2}\left [{\sin
k_S R \over k_S R} - \cos k_S R \right ]
\eqno (3.5a)$$
$$
[\xi_S(\vec{x}_1,t),\pi_L(\vec{x}_2,t)] =
a(t)^3
[\Phi_L(\vec{x}_1,t),\xi_M(\vec{x}_2,t)]
=
{i \over 2\pi^2} k_S^{~3}
H_0 {\sin k_S R \over k_S R}.
\eqno (3.5b)$$

Consider first the commutator between $\Phi_L$ and $\pi_L$.
Because it is a commutator between the coarse-grained objects
one would expect it to be finite when the two spacetime points are within
the same coarse-grained volume, and to vanish when they are far apart.
It is easy to work out either limit, but these results do not appear very
illuminating. Instead it would be better to use smeared fields. For example,
the canonical commutator can be
interpreted as
$$
\int_{V_y} d \vec{y}~ [\Phi (\vec{x},t), \pi (\vec{y},t)]
=i ~~~~~~~~~~~~x \in V_y   ,
$$
$$
{}~~~~~~~~~~~~~~~~~~~~~~~~~~~~~~=0 ~~~~~~~~~~~~x \notin V_y.
\eqno (3.6)$$
In flat spacetime the size of $V_y$ is not important. The interpretation is
just that $\Phi$ and $\pi$ are canonically conjugate variables if $\vec{x}$
and $\vec{y}$ are within the same smeared point inside the light cone.
Otherwise, $\Phi$ and $\pi$ can be measured independently. The smearing
can be done symmetrically, over both spatial arguments, but this just
produces an overall constant.

In inflationary universe it is natural to smear commutators over the
coarse-graining volume. Comoving coarse-graining scale is equal to $k_S^{-1}$.
Taking one of the spacetime points to be the origin, the coarse-grained
commutator in the first case turns out to be,
$$
\int_{V_S} d\vec{x}~ [\Phi_L(0,t),\pi_L(\vec{x},t)]=
{\cal O}(1)~i ~~.
\eqno (3.7a)$$
In the second case, one finds
$$
\int_{V_S} d\vec{x}~ [\Phi_L(0,t),\pi_L(\vec{R}_0+\vec{x})]\simeq -{2
i \over 3 \pi}\cos (k_SR_0)~(k_S R_0)^{-2}~~,
\eqno (3.7b)$$
where $R_0\gg k_S^{-1}$ is the mean comoving distance between the two smeared
volumes.

These relations show that on a given coarse-grained volume $\Phi_L$ and
$\pi_L$ do behave as canonically conjugate variables. The difference is that
unlike original field operators, the coarse-grained ones appear to have finite
commutator even outside the light cone. However the
commutator (3.7b) rapidly goes to zero, as $(k_S R_0)^{-2} \sim
(l/l_{phys}(R_0))^2 \sim \hbox{e}^{-2 H_0 t}$, where time is counted from the
moment when the two points left the same coarse-grained volume. That is, the
commutator becomes negligible as every point in one coarse-grained domain
gets out of the light cone of any point from the other coarse-grained domain.
According to Habbib [11], this decaying tail in the commutator (3.7b) is an
artefact of a sharp cut-off in the momentum space, Eq.'s (3.1-3.2).

The morale is that there is no magic just in writing down the coarse-grained
Klein-Gordon equation. At this level the theory
looks as quantum mechanical as it can be. The important simplification
is that instead of dealing with the full quantum field theory, we have
to deal only with the quantum mechanics in the Hilbert space defined by
$\Phi_L$ and $\pi_L$ as the two conjugate variables. (I will ignore here a
scaling factors implied by the need to turn commutator (3.7a) into precise
canonical form.) But it still looks as a quantum theory, and we still have
no argument to interpret Eq. (2.1) as Langevin equation.

It is however the commutator (3.5b) that contains the
germ of the eventual argument for the nearly classical nature of the
coarse-grained field.
Suppose that we go to the coordinate representation where $\Phi_L$ is
represented as a c-number and $\pi_L$ as derivative operator.
Then, the operator $\xi_S$ must have in that case a multiplicative, c-number
representation, while $\xi_m$ must have derivative piece.

Since total force term is the sum of the two, Eq. (3.3a), it is useful
to asses the relative strengths of the two contributions. The result is
remarkable. Let $t_{lc}(k)$ be the time at which the mode $k$ crosses the
coarse-graining scale $l$. That is, $a(t_{lc}(k)) k^{-1} = l$. The
relative magnitude of the two terms is given by the ratio,
$$
{\xi_S^{rms} \over \xi_m^{rms}} = {1 \over 3H_0} \left | {\dot {v}_k(t_{lc}(k))
\over v_k(t_{lc}(k))} \right |~~.
\eqno (3.8)$$
The explicit calculation must be done separately for cases of zero and finite
mass. The results are, respectively,
$$
{\xi_m^{rms} \over 3H_0} = {1 \over 3} \epsilon_S^2 \xi_S^{rms} ~~,
\eqno (3.9a)$$
$$
{\xi_m^{rms} \over 3H_0} = {1 \over 3} |\nu - 3/2| \xi_S^{rms} ~~,
\eqno (3.9b)$$
with,
$$
\nu^2 \equiv {9 \over 4} - {m^2 \over H_0^2} ~.
\eqno (3.9c)$$
These calculations are slight extension of those due to Sasaki, Nambu and
Nakao [12], and Hosoya, Nakano and Morikawa [12].
Starobinsky [6] already recognized that contribution due to
$\xi_m$ may be neglected in case of a massless scalar field.

It is therefore necessary to discuss three separate cases:

(i) Case of a zero mass. If coarse-graining radius $l$ is much larger than
the Hubble radius, the entire force term $\xi$ is well approximated by
the force term $\xi_S$. Thus, $\xi$ is represented by a $c$-number in the
coordinate representation.

(ii) Case of a small mass. If mass of the scalar field is small compared
to the Hubble parameter, so that $2m^2/(9H_0^2) \ll 1$, the contribution
of $\xi_m$ is suppressed for a factor of $m^2/(9H_0^2)$ compared to the
strength
of $\xi_S$. Again, $\xi \approx \xi_S$, and the same conclusion applies
about the $c$-number representation of the force term.

(iii) Case of a large mass. If mass of the scalar field is of the order of
Hubble parameter or larger, force term $\xi_m$ cannot be neglected
compared to $\xi_S$, and $\xi$ has significant derivative
piece in the coordinate representation. I do not consider that case
here in any significant detail.

The implications of these facts will be clearly seen in the Schroedinger
picture.

\vskip0.5truein

{\bf 4. Schroedinger picture}
\medskip
To develop Schroedinger picture for quantum field theory in any curved
spacetime it is most convenient to use the
rescaled field $\chi$: $\Phi(\eta, \vec {x}) \equiv \chi(\eta, \vec
{x})/S(\eta)$. In de Sitter case the scale factor expressed in the
conformal time is
$S(\eta) = - (H_0\eta)^{-1}$.

The Hamiltonian for the rescaled field is the same as for a system of
uncoupled oscillators in flat spacetime and with time dependent
frequency,
$$
\omega(\eta)^2 = S(\eta)^2 (m^2 - 2 H_0^2)~~.
\eqno (4.1)$$
The difference between the long modes, which make up the coarse-grained
field, and the short modes, whose inflow lead to the Langevine-like force
term, shows up now in the behavior of the effective one-mode frequencies,
$\omega_k(\eta)^2 \equiv k^2 + \omega(\eta)^2$. When mass is large compared
to the Hubble parameter all oscillators have real frequencies. However, for
$m^2 =0$, or $m^2 \ll H_0^2$, we have $\omega(\eta)^2 < 0$, and only short
modes, whose wavelengths at a given time are well within the Hubble radius,
have a real frequency $\omega_k(\eta)$. Modes
at longer wavelengths behave as amplitudes for upside down
harmonic oscillators. For a mode with a given wave number $k$, the time of
crossing the Hubble radius is roughly the time of transition from stable
oscillations to unstable growth.

This distinction has been noted by number of authors. In particular it has
been used by Guth and Pi [4] in their analysis of the quantum evolution of
the inflaton field. Here I attempt to develop this picture further, by
calculating and discussing the wave function for a coarse-grained field.
The key element is dynamical role of the continuous inflow of the short modes,
demonstrated so prominently in the Starobinsky's approach.

Bearing in mind this difference in physical picture, the
coarse-graining in the rescaled representation may be done in
a straightforward way, following Starobinsky [6].
The final coarse-grained field equation has a simple form:
$$
\chi_L^{\prime\prime}(\eta)  + \omega(\eta)^2\chi_L(\eta) =
\tilde{\xi}(\eta)~~.
\eqno (4.2)$$
The force terms $\tilde \xi$ arises as the difference between the
coarse-grained acceleration and the acceleration of the coarse-grained field,
but now it contains both force terms found before:
$$
\tilde{\xi}
\equiv
\langle \chi^{\prime\prime}\rangle_L - \chi_L^{\prime\prime}
=
S(\eta)^3(\xi_m + 3H_0\xi_S)~~.
\eqno (4.3)$$
All this is just algebra, and the result applies to the field of an arbitrary
mass. The physics is that of an oscillator that experiences both the
squeezing, due to the time dependent frequency, and the external force, due
to the inflow of the short modes. The effective Hamiltonian is:
$$
H_{eff}(\chi_L) = {1 \over 2} \left [ p_L^2 + \omega(\eta)^2\chi_L^2
\right ] - \tilde{\xi} \chi_L ~~,
\eqno (4.4)$$
where $p_L \equiv \chi_L^{\prime}$.

To write down the Schroedinger equation one has to recall the commutation
relations. Operators $\chi_L$ and $p_L$ make canonical pair, so they have
respectively a multiplicative and derivative representations.
The new moment is that while $\xi_S$ commutes with $\chi_L$ and does not
commute with $p_L$, the operator $\xi_m$ does not commute either with
$\chi_L$ or with $p_L$, as the latter contains $\Phi_L$ proportional piece.
The representation of the force term therefore significantly depends on
the mass of the scalar field.

The case of a large mass, where contribution of $\xi_m$ is dominant,
will be considered elsewhere. In cases with zero or small mass we have
$\tilde {\xi} \sim \xi_S$, and one finds the following effective
representation for the force term:
$$
\tilde {\xi} = \xi_c \cdot {\bf 1} + (-i)3H_0 S^2 D_2 \chi_L~~.
\eqno (4.5)$$
$D_2$ stands for the commutator between $\xi_S$ and $\pi_L$, while
$\xi_c$ is a c-number whose presence is allowed by the commutators.
Corresponding $c$-number piece in the representation of the force
term $\xi_S$ itself will be denoted with $\xi_{Sc}$, where $\xi_c =
3H_0 S(\eta)^3 \xi_{Sc}$. As we shall see, it is that term that is
responsible for all the interesting  physics, while the more elaborate
$D_2$ term in (4.5) presents merely a technical nuisance.
Upon substituting (4.5) into effective Hamiltonian (4.4) it is seen that
the only effect of the $D_2$ term is to introduce a shift in the oscillator
frequency,  $\omega^2 \rightarrow \omega_D^2 \equiv \omega^2 + (-i)
6H_0^2 S^2 D_2$. This shift becomes important only at very late times
and brings no surprises. Another way to arrive to the same conclusion would
be if instead of just neglecting $\xi_m$ we make a stronger assumption that
commutator $D_2$ vanish. The reason is that we seek representation of
the algebra (3.4-3.5) in the Hilbert space defined
by $\chi_L$ and $p_L$. By (3.5b) setting $\xi_m$ approximately to zero
implies vanishing $D_2$ as well. Therefore in what follows I will not
any more keep $D_2$ term in (4.5).

Thus, for the cases of a massless or nearly massless field, when $\tilde
{\xi}$ is represented by $\xi_c$, the Schroedinger equation has the following
form:
$$
 -{1 \over 2} \partial_{\chi_L \chi_L}^2 \Psi (\chi_L, \eta) +
\left [ {1 \over 2} \omega(\eta)^2 \chi_L^2 - \xi_c \chi_L \right ]
\Psi (\chi_L, \eta) = i \partial_{\eta}\Psi(\chi_L, \eta) ~~.
\eqno (4.6)$$

The physical nature of the solution to this equation is well known for some
special cases. If $\xi_c = 0$ and $\omega$ is a constant, we are dealing with
the linear harmonic oscillator and textbook solutions are either ground state
or coherent state, depending whether initial condition contains zero or
finite displacement. If $\omega$ is constant, we have forced linear harmonic
oscillator and solution is coherent state with the displacement due to the
action of the external force , see e.g. Ref. 13.
Finally, if $\xi_c = 0$, solution is a squeezed (vacuum) state, due to time
dependence in the frequency $\omega$ for this oscillator.

Since all these effects are contained in our case, the relevant solution
should have the following form:$^{14}$
$$
\Psi_{cs} (\chi_L, \eta)= A(\eta)~ \exp [ip_c(\eta) \chi_L] ~
\exp \left [-  {B(\eta) \over 2} [\chi_L - \chi_c(\eta) ]^2 \right ]~~.
\eqno (4.7)$$
This wave function is denoted with the subscript $cs$ to indicate that it is
both a coherent state and a squeezed state. Let us now work it out more
explicitly.

After substituting the ansatz (4.7) to the Schroedinger equation (4.6), and
after grouping the terms with the same powers of $\chi_L$, one obtains
three equations which allow us to find both the physical meaning and the
explicit solutions for parameters $A$, $B$, $\chi_c$, and $p_c$.
The results are as follows.

Parameter $\chi_c$ may be chosen to obey the classical equation of motion:
$$
\chi_c^{\prime \prime} + \omega^2(\eta)\chi_c = \xi~_c~.
\eqno (4.8)$$
Then, the parameter $p_c$ turns out to be momentum conjugate to $\chi_c$:
$p_c = \chi_c^{\prime}$.

The amplitude $A(\eta)$ may be written as,
$$
A(\eta) = A_0~\exp \left [-i S[\chi_c] \right ]~~
\exp \left [ -{i \over 2}\int_{\eta_0}^{\eta}
d\eta^{\prime}~ B(\eta^{\prime}) \right ]~~.
\eqno (4.9)$$
The first term, $A_0$, is a normalization constant, and its precise value is
of no interest here. Second term is the Feynman amplitude for classical
history $\chi_c(\eta^{\prime})$. The last term depends on the dispersion $B$
which is given as,
$$
B(\eta) = - i{\chi_h^{\prime} \over \chi_h}~~.
\eqno (4.10a)$$
Function $\chi_h$ is solution to the classical oscillator equation,
$$
\chi_h^{\prime\prime} + \omega^2(\eta) \chi_h =0~~,
\eqno (4.10b)$$
$$
\chi_h(\eta) = \alpha_+ (i\eta)^{p_+} + \alpha_- (i\eta)^{p_-}~~,
{}~~~~~~~
p_{\pm} = {1 \over 2} \pm \nu ~~.
\eqno (4.10c)$$

Once the wave function is known, everything may be rewritten in terms of
old variables $\Phi_L$ and $t$.
Using $\chi_L = a(t)\Phi_L$, and $\Phi_c(t) \equiv
\chi_c(\eta)/S(\eta)$, one finds,
$$
\Psi_{cs}(\Phi_L, t) =
A~\exp \left [ iS[\Phi_c]~ \right ]~
\exp \left [ - {i \over 2} \int_{t_0}^t dt^{\prime}~{B(t^{\prime}) \over
a(t^{\prime})} \right ]~~\times
$$
$$
{}~~~~~~~~~~~~~~~~~~~~~~~~~~~~~~~~~~~~~~\times
\exp [i p_c(t)a(t)\Phi_L]~
\exp \left [ - {B(t)a(t)^2 \over 2} \left (\Phi_L - \Phi_c(t) \right
)
\right ]~~.
\eqno (4.11)$$

The width of that wave function is,
$$
\sigma_{\Psi}^2 =
[a^2(t)(B(t) + B^*(t))]^{-1} ~~.
\eqno (4.12)$$

This completes the discussion of the structure of the wave function. The main
conclusion is that in cases of a massless  field, or field with a finite, but
small mass, the amplitude for a coarse-grained field is both a coherent
and a squeezed state. In next two Sections it will be seen how that works.

\vskip0.5truein

{\bf 5. Case of a zero mass}
\bigskip

I will first consider the simplest and much discussed case of a massless
field. The form of the wave function is already known from the preceding
section, and one just has to figure out explicit expressions for $\nu = 3/2$.
The powers in solution (4.10d) become $p_+=2$, and $p_- = -1$. Consequently,
$$
B(\eta) = - {i \over \eta}
{{2 i \alpha_+ \eta^3 + \alpha_-} \over
{i\alpha_+\eta^3 - \alpha_-}} ~~;
\eqno (5.1a)$$
$$
B + B^* = {{(-)3\eta^2(\alpha_+\alpha_-^* + \alpha_+^*\alpha_-)} \over
{|\alpha_+|^2\eta^6 + (-i)(\alpha_+\alpha_-^* - \alpha_+^*\alpha_-)\eta^3 +
|\alpha_-|^2}}~~.
\eqno (5.1b)$$

At ``late times,'' that is, in the limit $\eta \rightarrow 0^-$,
{\cal Re}$B$ approaches asymptotic form, $B + B^* = C \eta^2 =
C~H_0^{-2}a(t)^{-2}$, where $C \equiv (-)3 (\alpha_+ \alpha_-^* +
\alpha_+^* \alpha_-)/|\alpha_-|^2$ is constant that depends on the initial
conditions. The onset of that asymptotics does depend on the actual values
of $\alpha_{\pm}$, but it always takes place, except when both
$\alpha_{\pm}=0$, in which case $B$ vanishes as well.
If {\cal Re}$(\alpha_+ \alpha_-^*) < 0$, {\cal Re}$B$ is strictly positive.

Let us now look at what coherent-squeezed wave function predicts about the
physics of a coarse-grained massless scalar field. From Eq.(4.11) it
follows,
$$
\langle \Psi_{cs}(t)|\Phi_L|\Psi_{cs}(t)\rangle = \Phi_c(t)~~,
\eqno (5.2a)$$
$$
\langle \Psi_{cs}(t)|~\Phi_L^2~|\Psi_{cs}(t) \rangle =
{H_0^2 \over C} + \Phi_c^2(t) ~~,
\eqno (5.2b)$$
$$
\ddot{\Phi}_c + 3H_0\Phi_c = 3H_0 \xi_{Sc} ~~.
\eqno (5.2c)$$

These expressions ought to be compared with the vacuum expectation values
calculated in the Heisenberg picture:
$$
\langle 0|\Phi_L(t)|0\rangle = 0~~,
\eqno (5.3a)$$
$$
\langle 0 |\Phi_L(t)^2 |0 \rangle = {H_0^3 \over 4\pi^2}(t - t_i)~~.
\eqno (5.3b)$$

If, on the basis of (5.2a) vs. (5.3a), one sets $\Phi_c(t) = 0$, there is no
hope to reproduce from (5.2b) linear in time growth of dispersion which is
predicted by (5.3b). Moreover, $\Phi_c$ may have a finite value, either due to
the initial conditions, or due to force term $\xi_{Sc}$. The later
represents a nontrivial operator $\xi_S$ which cannot be
identically equal to zero.

This is the crucial point in the analysis. As seen from Eq. (5.2b), the width
of the wave function is constant in time. In order for Schroedinger
picture to reproduce the well known results from the Heisenberg picture,
the relevant contribution has to come
from $\xi_{Sc}$ generated contribution to parameter $\Phi_c(t)$. For this,
the following two sufficient conditions have to be met:

(i) In coordinate Schroedinger representation, force term $\xi_S$ has to be
represented by a $c$-number valued classical random function $\xi_{Sc}(t)$
whose expectation values mirror the vacuum expectation values of the quantum
operator $\xi_S(t)$ in the Heisenberg representation:
$$
\langle \xi_{Sc} (t)\rangle_c = 0 ~~;
\eqno (5.4a)$$
$$
\langle \xi_{Sc}(t_1)~\xi_{Sc} (t_2)\rangle_c = {H_0^3 \over 4\pi^2}
\delta (t_1 - t_2) ~~,
\eqno (5.4b)$$
where $\langle ~\rangle_c$ stands for the average over $\xi_{Sc}$;

(ii) the expectation values in the Schroedinger picture have to be
computed in two steps: first, with respect to the given state $\Psi_{sc}$,
second, with respect to the classical random variable $\xi_{Sc}$.

The ``proof'' for these rules is easy, as the correspondence between the two
pictures is now restored:
$$
\langle 0 | \Phi_L(t) \ 0 \rangle = \langle ~
\langle \Psi_{sc}(t)| \Phi_L | \Psi_{sc}(t) \rangle ~\rangle_c ~~,
\eqno (5.5a)$$
$$
\langle 0 | \Phi_L^2(t) \ 0 \rangle = \langle ~
\langle \Psi_{sc}(t)| \Phi_L^2 | \Psi_{sc}(t) \rangle ~\rangle_c ~~.
\eqno (5.5b)$$

The legitimate question at this moment is whether all this mean anything?
There are several papers, for instance, Brandenberger in Ref. 5,
where some classical random variable has been introduced, citing this
step as ``inevitable in any kind of semiclassical analysis.'' And many more
authors made similar steps silently. However, in all those cases, assumptions
has always been made that some kind of a split to ``quantum'' and
``classical'' components of the scalar field $\Phi$ can be made, and that the
``classical'' part exists. No such assumption has been made here. The
necessity to introduce classical random variable $\xi_{Sc}$ comes from the
need to develop the Schroedinger picture that would agree with the Heisenberg
picture, that is, by staying fully within the rules of the quantum theory
without assuming any classical realm whatsoever. The necessity to introduce
classical random variable $\xi_{Sc}$ is no more peculiar than the necessity to
introduce the derivative representation for the canonical variable $\pi_L$.

Another point worth mentioning is that while (5.2a) and (5.3a) are now
reconciled automatically due to (5.5a), Eq. (5.5b) is true only up to the
corrections due to the width of the coherent-squeezed state in
(5.2b). This striking feature has following implication.

Equation (5.2c) is exactly solvable and it is easy to see that for $t \gg
H_0^{-1}$ the second derivative term may be neglected with respect to the
other two. The position of the mean of the wave function is accurately
described at late times by the classical equation,
$$
\dot {\Phi}_c(t) = \xi_{Sc} ~~.
\eqno (5.6)$$
This is sought for Langevin equation, the true classical
version of the Starobinsky equation (2.1) for the massless case.
{}From Eq.'s(5.4a,b) it follows that,
$$
\langle \Phi_c^2(t)\rangle_c = \Phi_c^2(0) + {H_0^3 \over 4\pi^2}t~~,
\eqno (5.7)$$
which is the true meaning of the well known Bunch-Davis result. One can now
compare at late times the width of the wave function with the Brownian spread
in the position of its mean to find,
$$
{\sigma_{\Psi}^2 \over {\langle \Phi_c^2\rangle_c}} \sim {4\pi^2 \over C(H_0
t)}~~.
\eqno (5.8)$$
Therefore, the width of the wave function
becomes negligible compared to the (classical) r.m.s. value of its mean.
In that sense the coarse-grained massless field is nearly classical.

\vskip0.5truein

{\bf 6. Case of a small mass}
\bigskip
In this case solutions to the classical oscillator equation (4.10c,d) have
powers, $p_+= 2 - m^2 /3H_0^2$, and, $p_-= -1 + m^2 /3H_0^2$. One finds,
$$
B = (-i){{|\alpha_+|^2 p_+ \eta^{2p_+ -1} + |\alpha_-|^2 p_- \eta^{ p_- -1}
+ (\alpha_+ \alpha_-^* p_+ + \alpha_+^* \alpha_- p_-) \eta^{p_+ + p_- -1}}
\over {|\alpha_+ \eta^{p_+} + \alpha_- \eta^{p_-} |^2}}
\eqno (6.1a)$$
$$
B + B^* = {{i(\alpha_+^*\alpha_- - \alpha_+ \alpha_-^*)(3 - 2m^2/(3H_)^2))}
\over {|\alpha_+ \eta^{2 - m^2/(3H_0^2)} + \alpha_- \eta^{-1 + m^2/(3H_0^2)}
|^2}} ~~.
\eqno (6.1b)$$
The later quantity is positive if $i(\alpha_+^*\alpha_- - \alpha_+\alpha_-^*)
> 0$. At late times the width of the wave function behaves as,
$$
[a^2(B+B^*)]^{-1} \rightarrow {{|\alpha_-|^2} \over {i(\alpha_+^*\alpha_- -
\alpha_+\alpha_-^*)}} {H_0^2 \over 3} \left (1 - {2m^2 \over 9H_0^2} \right )
|\eta|^{2m^2/(3H_0^2)}~~.
\eqno (6.2)$$

Let us again compare the expectation values computed in a
coherent-squeezed quantum state with vacuum expectation values in the
Heisenberg picture. Respectively, they are,
$$
\langle \Psi_{cs}(t)|\Phi_L|\Psi_{cs}(t)\rangle = \Phi_c(t)~~,
\eqno (6.3a)$$
$$
\langle \Psi_{cs}(t)|~\Phi_L^2~|\Psi_{cs}(t) \rangle =
C H_0^2 \exp \left [- {2m^2 \over 3H_0} t \right]
+ \Phi_c^2(t) ~~,
\eqno (6.3b)$$
$$
\ddot{\Phi}_c + 3H_0 \dot {\Phi}_c + m^2 \Phi_c = 3H_0 \xi_{Sc} ~~,
\eqno (6.3c)$$
and,
$$
\langle 0|\Phi_L(t)|0\rangle = 0~~,
\eqno (6.4a)$$
$$
\langle 0 |\Phi_L(t)^2 |0 \rangle = {3 H_0^4 \over 8\pi^2 m^2}~~.
\eqno (6.4b)$$
In Eq. (6.3b) above $C$ stands for the dimensionless constant whose definition
is apparent from (6.2). It depends on the initial conditions and vanishes
only when $\alpha_- = 0$.

One again concludes that the key for the reconciliation of the two
pictures is in the representation of the force term. Let us make the same
two assumptions as in Section 5. Taking $\langle ~ \rangle_c$ average of
Eq. (6.3a) immediately yields agreement with (6.4a), assuming the same
initial conditions, of course. To establish the same for (6.3b) vs. (6.4b),
one has to look more closely into solution to the classical equation of
motion for the center of the quantum state, Eq. (6.3c). With retarded
Green function the solution is,
$$
\Phi_c(t) = c \exp \left [ {-m^2 \over 3H_0} t \right ] +
d \exp [-3H_0 t]
- \int_0^t dt^{\prime}~\left [
\exp [3 H_0 (t^{\prime} - t)] +
\exp \left [{m^2 \over 3H_0} (t^{\prime} - t) \right ] \right ]
\xi_{Sc}(t^{\prime}) ~~.
\eqno (6.5)$$
Consider now Eq. (6.3c) but without the second derivative term:
$$
\dot{\Phi}_c(t) = - {m^2 \over 3H_0} + \xi_{Sc} ~~.
\eqno (6.6a)$$
The solution is,
$$
\Phi_c(t) = \Phi_c(0) \exp \left [{- m^2 \over 3H_0} t \right ] +
\int_0^t dt^{\prime}~
\exp \left [{m^2 \over 3H_0} (t^{\prime} -t) \right ]
\xi_{Sc} (t^{\prime}) ~~.
\eqno (6.6b)$$
The two $\xi_{Sc}$-independent terms in (6.5) and (6.6b) agree at
$H_0t \gg 1$. The same is true for the force dependent terms, as may be seen
by comparing the dispersions. Exact solution (6.5) yields,
$$
\langle \Phi_c^2(t)\rangle_c = \Phi_c^2(0)
\exp \left [{-2 m^2 \over 3H_0} t \right ] +
{{3H_0^4} \over {8 \pi^2 m^2}} \left ( 1 -
\exp \left [{-2 m^2 \over 3H_0} t \right ] \right )
$$
$$
{}~~~~~~~~~+ {{3H_0^4} \over {2 \pi^2(H_0^2 + m^2)}}\left ( 1 -
\exp [- 3H_0(1 + m^2/(9H_0^2)) t] \right )  +
{{H_0^2} \over {24 \pi^2}}\left ( 1 -
\exp [- 6 H_0 t] \right )~~ ,
\eqno (6.7a)$$
while solution (6.6b) leads to,
$$
\langle \Phi_c^2(t) \rangle_c =\Phi_c^{~2}(0)
\exp \left [{-2 m^2 \over 3H_0} t \right ] +
{{3H_0^{~4}} \over {8 \pi^2 m^2}} \left ( 1 -
\exp \left [{-2 m^2 \over 3H_0}t \right ] \right )~~.
\eqno (6.7b)$$
The two experessions agree at times large compared to the Hubble time, and up
to the corrections of the order of $m^2/H_0^2$.

To summarize, in case of a small mass the reconciliation between the two
pictures is also achieved if the force term $\xi_S$ is represented by the
$c$-number valued classical random function $\xi_{Sc}$. The effective
equation of motion for the position of the mean of the
wave function is a true Langevin equation (6.6a). This is the proper
classical version of the Starobinsky equation (2.1). $\langle~\rangle_c$
averages of the wave function expectation values reproduce the vacuum
expectation values found in the Heisenberg picture. The relaxation factor
in Eq.'s (6.7a,b) has been found before by mode summation [7]
as a correction to the Bunch-Davies result in case of inflationary universe
of a finite duration.

Comparing the width of the wave function with the stochastic spread of
its mean yields at late times,
$$
{\sigma_{\Psi}^2 \over {\langle \Phi_c^2 \rangle_c}} \simeq
{8 \pi^2 C \over 3}
{m^2 \over H_0^2} \exp \left [ (-){2m^2 \over 3H_0} t \right ]~~.
\eqno (6.8)$$
The characteristic time in the exponent is the duration of a classical
inflationary phase driven by the $m^2 \Phi^2$ potential. There is also
a suppression due to the assumed small value of the mass compared to the
Hubble parameter. Thus, one concludes that in the case of a field with a small
mass, the quantum state of a coarse-grained field has shrinking width
with respect to the r.m.s. value for the position of its mean, $\langle
\Phi_c^2 \rangle_c^{1/2}$. The evolution is effectively classical.

\vskip0.5truein
{\bf 7. Conclusion}
\bigskip
The goal of this work was to establish the Langevin interpretation for
coarse-grained quantum field equation (2.1), and thereby establish a
highly classical and stochastic nature of the coarse-grained quantum field
in de Sitter space. This has been accomplished by constructing the
Schroedinger representation and calculating the wave function for the
coarse-grained field, which turns out to be both a coherent and a squeezed
state. The center of that wave function effectively obeys classical equation
(6.6a) which parallels the original coarse-grained quantum field equation. By
calculating in the Schroedinger picture the vacuum expectation values well
known from the Heisenberg picture, one may deduce the Schroedinger picture
representation for the force term which models the influence of the inflow of
the short modes on the coarse grained field. It turns out that for a massless
or
low mass fields, this quantity is represented by a classical random noise.

This has two immediate consequences.
One, that the vacuum expectation value in the Heisenberg
picture corresponds to the double average in the Schroedinger picture, first
with respect to the width of the wave function, then with respect to the
Brownian walk of its mean: $\langle 0 |{\cal O}(t)|0\rangle =
\langle ~ \langle \Psi_{cs}(t)| {\cal O} |\Psi_{cs}(t) \rangle ~\rangle_c$.
Two, the effective equation for the evolution of the center of the wave
function, Eq. (6.6a), is a true Langevin equation.

Moreover, one finds that at times larger than the Hubble time, the width of
the quantum state becomes negligible compared to the r.m.s. of its mean due to
the Brownian spread. Therefore, to a high accuracy, the quantum state of a
coarse-grained field obeys the clustering property,
$$
\langle \Psi_{sc}(t)| \Phi_L^2 |\Psi_{sc}(t) \rangle
\approx \langle \Psi_{sc}(t)| \Phi_L |\Psi_{sc}(t) \rangle ^2~~,
\eqno (7.1)$$
in a sense that $\langle ... \rangle_c$ of both sides agree at late times
up to corrections due to the small, finite width of the wave function.
Quantum evolution may be pictured as a random walk of almost-$\delta$-function
on a bundle of classical trajectories. The only
evolution worth keeping track of is that of the mean value which evolves
according to the Langevin equation (6.6a). We firmly arrived at the
Starobinsky's picture. One may now proceed with either analyzing the solutions
to the Langevin equation, or by writing down appropriate Fokker-Planck
equation for the distribution over the mean value $\Phi_c$.
Both methods have been applied to various
inflationary models and different boundary and initial conditions.

With respect to the density perturbations, the original analysis of Guth and
Pi [4] appears more relevant for the traditional treatment [2] when one does
mode by mode decomposition, while coarse-grained picture may be useful for
the approach of Ellis and Bruni [15] where a priori one does not have to do
the mode expansion. The relevance of the squeezed states for the generation of
density perturbations, again in a mode-by-mode treatment, have been recognized
for a long time by Grishchuk [16].

Potentially the most useful methodological outcome of this work might
be in further investigations of the meaning and the interpretation of
wave functions in quantum cosmology. Starting again with Starobinsky's paper
[6], there is a definite, positive identification of the square of the
minisuperspace wave function with the solutions to the Fokker-Planck equation
in stochastic inflation [17, 9, 18]. The physics behind the
emergence of classical properties in models considered here, should clearly
help us to understand better the decoherence in quantum cosmology.

This discussion also underlies a seldom appreciated fact, that it is not
necessary to have some interaction in order to have the emergence of a
highly classical behavior in quantum systems. The models considered here
were free fields, but of course, the point is that they are not quite free:
the background is nontrivial. In rescaled representation the effect of the
background shows as a particular time dependence of the oscillator frequency.
The effect of a small or vanishing mass is to turn this oscillator into
upside down one. And the effect of coarse-graining is to provide an
external force acting on the oscillator which describes coarse-grained field.
All three effects together lead to the classical behavior. It is apparent
that the same conclusion applies to any other physical system whose
description may be given in terms of such kind of oscillator. It is also
apparent that these features alone suffice to make any long wavelength mode
in inflationary universe effectively classical. Other arguments, like the
decoherence due to the coupling between the modes, come on top of the argument
described here, and provide another source for nearly classical
behavior.

\vfil\eject
{\bf Acknowledgment}

During part of the work reported here I had a benefit of many close
discussions with Salman Habib which are all pleasure to acknowledge. His
enthusiastic readiness to examine every angle in detail were very
stimulating and are much appreciated.

\vskip0.5truein


\def\tindent#1{\indent\llap{#1}\ignorespaces}
\def\ref{\par\hang\tindent}

{\bf References:}
\medskip
\ref{$1.$ }S.W. Hawking, {\it Phys. Rev.} {\bf D37}, 904, (1988);

\ref{$1.$ }A. H. Guth, {\it Phys. Rev.}, {\bf D23}, 347, (1981);
for reviews on inflation, see L. F. Abbott and S.-Y. Pi (eds.), {\sl
Inflationary Cosmology}, (World Scientific, Singapore, 1986); A. D.
Linde, {\sl Particle Physics and Inflationary Cosmology}, (Harwood
Academic Publishers, New York, 1990); and references therein.
\ref{$2.$ }S. W. Hawking, {\it Phys. Lett.}, {\bf 115B}, 295 (1982);
A. A. Starobinsky, {\it Phys. Lett.}, {\bf 117B}, 175 (1982);
A. H. Guth and S.-Y. Pi, {\it Phys. Rev. Lett.}, {\bf 49}, 1110 (1982);
J. M. Bardeen, P. J. Steinhardt, and M. S. Turner, {\it Phys. Rev.},
{\bf D28}, 679 (1983).
\ref{$3.$ }A. D. Linde, {\it Mod. Phys. Lett.}, {\bf 1A}, 81 (1986);
{\it Phys. Lett.}, {\bf 175B}, 395 (1986);
A. S. Goncharov, A. D. Linde, and V. F. Mukhanov,
{\it Int. J. Mod. Phys.}, {\bf A2}, 561 (1987).
\ref{$4.$ }A. Guth and S. -Y. Pi, {\it Phys. Rev.}, {\bf D32}, 1899, (1985).
\ref{$5.$ }A.A. Starobinsky, Ref. 2; R. H. Brandenberger, {\it Nucl. Phys.},
{\bf B245}, 328 (1984);
R.H. Brandenberger, R. Laflamme, and M. Miji\'c, {\it Mod. Phys. Lett.},
{\bf A28}, 2311, (1990); see also Ref. 12.
\ref{$6.$ }A. A. Starobinsky, in {\sl Fundamental Interactions}
(MGPI Press, Moscow, 1983); in {\sl Current
Topics in Field Theory, Quantum Gravity, and Strings} edited by H. J.
deVega and N. Sanchez (Springer, New York, 1986).
\ref{$7.$ }A. Vilenkin and L. H. Ford, Phys. Rev. D {\bf 26},
1231 (1982); A. D. Linde, Phys. Lett. {\bf 116B}, 335 (1982); A. A.
Starobinsky, Ref 2; A.D. Linde in {\sl The Very Early Universe}, ed. by
G.W. Gibbons, S.W. Hawking, and S.T.C. Siklos, (Cambridge University
Press, 1983).
\ref{$8.$ }A. Vilenkin, {\it Phys. Rev.}, {\bf D27}, 2848, (1983).
\ref{$9.$ }M. Miji\'c, {\it Phys. Rev.}, {\bf D42}, 2469 (1990).
\ref{$10.$ }M. Gell-Mann and J.B. Hartle, in {\sl Proc. of 3rd Int.'l
Symp. on the Found. of Quant. Mech. in Light of New Technology}, ed. by
S. Kobayashi et al., (Physical Society of Japan, Tokyo, 1990).
\ref{$11.$ }S. Habib, (unpublished).
\ref{$12.$ }A. Hosoya, M. Morikawa and K. Nakayama, {\it Int. J. Mod. Phys.},
{\bf A4}, 2613, (1989);
M. Sasaki, Y. Nambu, and K. Nakao, {\it Nucl. Phys.}, {\bf B308}, 868, (1988).
\ref{$13.$ }P. Carruthers and M. M. Nieto, {\it Am. J. Phys.}, {\bf
33}, 537 (1965).
\ref{$14.$ }M. Miji\'c, talk at the Tufts University
{\sl Symposium on Quantum Cosmology},
May 1989 (unpublished).
\ref{$15.$ }G.F.R. Ellis and M. Bruni, {\it Phys. Rev.}, {\bf D40}, 1804,
(1989).
\ref{$16.$ }L.P. Grishchuk, {\it Phys. Rev.}, {\bf D48}, 5581,
(1994); and references therein.
\ref{$17.$ }A. S. Goncharov and A. D. Linde, {\it Sov. J. Part. Nucl.},
{\bf 18}, 369 (1986).
\ref{$18.$ }M. Miji\'c, {\it Int. J. Mod. Phys.}, {\bf A6}, 2685, (1991).

\end